# Neutrinos and Asteroseismology of Stars over the Helium Flash

Diogo Capelo and Ilídio Lopes
CENTRA—Centro de Astrofísica e Gravitação, Instituto Superior Técnico, Universidade de Lisboa, Av. Rovisco Pais 1, 1049-001 Lisboa, Portugal
diogo.capelo@tecnico.ulisboa.pt, ilidio.lopes@tecnico.ulisboa.pt
Received 2022 December 19; revised 2023 June 16; accepted 2023 June 30; published 2023 August 16

## Abstract

The helium flash, occurring in stars of 0.6–2.0 $M_\odot$ at the end of the red giant branch, is not observable via optical means due to the energy of the process being used to lift the core out of degeneracy. Neutrinos, which are linked to the ignition of reactions triggered during the flash and serve as the only cooling process in the inert core, can help characterize changes in internal structure. In this work, we create 18 stellar models across three mass and six metallicity values, chosen in the context of the stellar abundance problem, to compare the evolutionary path up to and probe the helium flash by conducting a detailed study of neutrino emission throughout this crucial phase of stellar evolution. We demonstrate how thermal neutrino emissions could have an imprint on global asteroseismic parameters and use them as an additional tool to infer the impact of compositional changes. We find that a precision of 0.3 $\mu$Hz in the determination of $\Delta \nu$ is enough to distinguish between between the two most prominent solar composition models and confirm that asteroseismic observation can be enough to classify a star as undergoing the process of helium subflashes. We also predict nuclear neutrino emission fluxes and their evolution for all relevant sources.

*Unified Astronomy Thesaurus concepts:* Asteroseismology (73); Neutrino astronomy (1100); Red giant tip (1371); Stellar abundances (1577)

## 1. Introduction

As stars burn hydrogen in their cores during the main sequence, they accumulate in their central regions the product of those reactions, helium, which becomes increasingly abundant as the star ages and the hydrogen in the core is depleted. Eventually, hydrogen burning moves to a shell around the core, and the star leaves the main sequence. For stars with masses $0.6\,M_\odot < M_* < 2.0\,M_\odot$, the conditions at the center are not extreme enough to ignite the fusion of the (mostly) helium core right away, and the inert nucleus is supported against gravity by the pressure of degenerate electrons (Sweigart & Gross 1978). This means that the mass of the degenerate core continues to increase, as does its temperature, as the hydrogen-burning shell moves farther and farther away from the center of the star, leaving more helium behind. Due to the characteristics of degenerate matter and a narrow temperature range for the onset of helium burning, around $10^8$ K, the conditions under which these stars rapidly ignite the He in the core (a process called the helium flash) can be well defined (Schwarzschild & Härm 1962).

Although predicted almost 60 yr ago, this phenomenon has never been observed, owing in part to the timescale of the process and in part to the fact that the thermal expansion of the star's previously degenerate core absorbs much of the released energy (Mocák et al. 2008, 2009). This makes detecting the helium flash difficult without some form of access to the star's central region, a difficult zone to probe using the traditionally more readily available tools. In the past few years, however, the rapidly growing fields of helio and asteroseismology, through the analysis of observable photometric variations in the surfaces of stars due to convection-driven pulsations, have succeeded in developing methods to characterize internal solar (Gough 1996; Basu et al. 2009; Chaplin & Miglio 2013) and stellar (Hekker & Christensen-Dalsgaard 2017; Lopes et al. 2019) structure, respectively, and test new (Gomes & Lopes 2020) and existing (Aerts 2021) physics. Additionally, the PLATO mission (Rauer et al. 2014; Miglio et al. 2017), due to be launched in 2026, is expected to both expand the current catalog of stellar oscillation observations and improve upon their already considerable precision (Goupil 2017).

Another method to study the interiors of stars would be to use neutrinos (Lopes & Turck-Chiéze 2013; Lopes 2013), as they are not affected by the outer layers of the star from the moment of their creation in nuclear reactions. The detection of (nuclear) solar neutrinos was, in fact, instrumental in resolving the neutrino oscillation problem (Ahmad et al. 2001), a long-standing problem in astrophysics. More recently, the first neutrinos from the CNO reaction occurring in the Sun were measured in the Borexino experiment (Borexino Collaboration 2020), confirming models for fusion inside the Sun and providing the first clues toward a solution of the abundance problem. More recently, the Borexino experiment's Phase III preliminary results (Kumaran et al. 2021) still validate both high- and low-metallicity compositions but show a slight skew toward the high-$z$ hypothesis. Though progress is being made, detecting the predominantly low-energy neutrinos emitted in the course of nuclear fusion from extrasolar sources remains out of reach to this day, as it would take extreme sources of emission, or one relatively nearby, for detection to be possible. This type of direct detection is therefore possible at this time only for very specific sources and types of emission, namely, Type II supernovae (Hirata et al. 1988) and active galactic nuclei (IceCube Collaboration et al. 2022). However, we provide our results with the expectation that an estimation of the currently predicted fluxes can help dimension future endeavors in this area.

With respect to the helium flash in particular, Bildsten et al. (2012) showed that global asteroseismic parameters, in conjunction with fundamental evolutionary information like mass, should be enough to classify a star as undergoing the helium flash process. In this work, we aim to expand on this result by testing the stability of these helium flash tracks with metallicity and composition changes. Apart from the specific challenges of understanding the composition of stars, this is of great importance to astrophysical research in general because stellar models, which are widely used to fit and draw conclusions from astrophysical data, depend strongly on the initially considered metallicity, a parameter that is very difficult to estimate and is affected by considerable uncertainties, even in the case of the Sun (Turck-Chiéze et al. 2004; Basu et al. 2012; Stancliffe et al. 2016; Tayar et al. 2022). This acquires added relevance because, without better constraints on the allowed stellar metallicities, while our observations of stars will continue to improve, our ability to model them will remain severely limited by uncertainties in fundamental quantities like metallicity, composition, and internal stellar processes like rotation and opacity (Turck-Chiéze et al. 2016; Yang 2019). We also examine the relation of neutrinos to the helium flash and demonstrate that asteroseismic observables can serve as an indirect measurement of their emission. Furthermore, we provide estimations of neutrino flux for relevant nuclear sources, including those from the flash burning of nitrogen in the core (Serenelli & Fukugita 2005), as a benchmark for an avenue of direct observation of the helium flash.

## 2. Model Building

Release version 12115 of the stellar evolution code MESA (Paxton et al. 2011; Paxton et al. 2013; Paxton et al. 2015, 2018; Paxton et al. 2019) was used to generate models from the pre–main sequence, assuming the stars were initially chemically homogeneous and fully convective. MESA already allows users to easily extract information regarding neutrinos according to the general process that originates them (fusion pp or CNO neutrinos and four different nonnuclear processes; e.g., Farag et al. 2020) but is not yet capable of specifying their source by reaction or the location in the star where they were produced. In order to obtain this information, it was necessary to modify the evolution code used to obtain the models. These custom modifications allow the computation of radial profiles of neutrino emission from the nuclear reactions pertaining to the pp chain and CNO cycle (e.g., Lopes & Turck-Chiéze 2013; Capelo & Lopes 2020) and also of the total source-dependent neutrino emission for the entire evolutionary history of the star.

Knowledge of nuclear reaction rates is of great importance to the precision of stellar models, as was noted (for the case of the Sun) by Villante & Serenelli (2021). MESA's nuclear reaction rates were taken from the (updated) JINA REACLIB (Cyburt et al. 2010), with the exception of the $^9\text{Be}(\alpha,n)^{12}\text{C}$ reaction, which is taken from Kunz et al. (1996).

Atomic opacities are another major factor affecting the models, and recent developments have shown that they may not be as well understood as is desirable. Work by Bailey et al. (2015) pertaining to the opacity of Fe, which contributes a significant part of the total opacity at the radiative/convective boundary, reveals discrepancies in both the quasicontinuum part and the peak part of the opacity spectrum, with an increase of 30%–400% relative to current theoretical predictions. Similar studies confirm the results obtained for Fe, but the results for other metals (namely, Cr and Ni) do not show a significant difference (Nagayama et al. 2019), further evidencing the need to refine the current understanding of opacity. Additionally, it has been found that a maximum increase of about 22% in opacity at the base of the solar convective zone (Bahcall et al. 2005; Christensen-Dalsgaard & Houdek 2010) would be enough to reconcile the results of high- and low-metallicity models and bring the latter into agreement with helioseismology (Christensen-Dalsgaard et al. 2009), providing a possible start to the most direct solution of the solar abundance problem, with a possible extension to the modeling of other stars. This would not be the first time a discrepancy between model predictions and observations was reconciled by improving opacity standards, as previous work on Cepheids demonstrates (Petersen 1973; Moskalik et al. 1992).

**Table 1**
Table of Model Metallicities

| Model  | Metallicity |
|--------|-------------|
| AG09−  | 0.01153     |
| AG09   | 0.01340     |
| AG09+  | 0.01445     |
| GS98−  | 0.01465     |
| GS98   | 0.01690     |
| GS98+  | 0.01834     |

In this work, the atomic opacities used are from the Opacity Project (Badnell et al. 2005), corrected for low temperatures with molecular opacities from Ferguson et al. (2005).

Using this same baseline, a set of models was computed. The set varies the mass (1.0, 1.3, and 1.6 $M_\odot$), the relative element abundance of the models (Grevesse & Sauval 1998, hereafter GS98; Asplund et al. 2009, hereafter AG09), and the metallicity, resulting in a total of 18 models. For each mass value, the models are designated by the name of their relative element abundance distribution (GS98 or AG09) with a plus or minus sign at the end indicating a modified higher or lower metallicity relative to the solar case, respectively. The metallicities of the models (equal for each value of mass) can be found in Table 1. Note that models AG09+ and GS98− differ in metallicity from each other by only 0.0002, or 1.37%. These are the most similar among the set of models.

A listing of the logarithmic abundances for some selected metals for each of the element abundance distributions considered in this work can be found in Table 2.

These models were tested by analyzing well-known characteristics of stars near and at the helium flash, namely, their fundamental properties and the stability of the tip of the red giant branch (TRGB). These results can be seen in Appendix A.

## 3. Asteroseismology

Stellar interiors can be characterized by analyzing two (global) asteroseismic parameters that relate asymptotically to quantities that are relevant to the physical processes occurring there. Oscillations exhibit two main behaviors. Modes for which pressure acts as the restoring force (p-modes) allow for the extraction of information regarding the stellar envelope. The large frequency separation measures the difference in the frequency between consecutive modes of the same angular degree, $\Delta\nu \approx \nu_{n,\ell} - \nu_{n-1,\ell}$, with $\nu = \omega/2\pi$, and is shown to relate to the mean density (and therefore the mass and radius)

Table 2
Logarithmic Abundances $\log \epsilon_i \equiv \log N_i/N_H + 12$ of the Most Relevant Metals for the Two Different Composition Determinations

| Element | GS98 | AG09 |
| --- | --- | --- |
| C | 8.52 | 8.43 |
| N | 7.92 | 7.83 |
| O | 8.83 | 8.69 |
| Ne | 8.08 | 7.93 |
| Mg | 7.58 | 7.60 |
| Si | 7.55 | 7.51 |
| S | 7.33 | 7.13 |
| Ar | 6.40 | 6.40 |
| Fe | 7.50 | 7.50 |
| $(Z_s/X_s)_\odot$ | 0.0231 | 0.0181 |

**Note.** Here $N_i$ is the number abundance of the element $i$.

of a star (Chaplin & Miglio 2013), depending on the speed of sound in the stellar interior (Tassoul 1980),

$$\Delta\nu = \left(2\int_0^R \frac{dr}{c(r)}\right)^{-1},\quad (1)$$

where $R$ is the total radius of the star, and $c(r)$ is the speed of sound at radius $r$.

The more elusive $g$-modes (for which the restoring force is gravity) propagate in the inner core and are therefore more difficult to observe directly. They can be described by their separation in period, $\Delta\Pi_l = \Pi_{n,\ell} - \Pi_{n-1,\ell}$, with $\Pi = 1/\nu$, which relates to the size of this region (Montalbán et al. 2013) and depends on the Brunt–Väisälä (or buoyancy) frequency (Tassoul 1980),

$$\Delta\Pi_\ell = \frac{2\pi^2}{\sqrt{l(l+1)}}\left(\int_{r_1}^{r_2} N\frac{dr}{r}\right)^{-1},\quad (2)$$

where $r_1$ and $r_2$ are the turning points of the $g$-mode cavity, i.e., the limits of the region of the star where the $g$-modes propagate; $\ell$ is the angular degree; and $N$ is the Brunt–Väisälä frequency. For dipole modes ($\ell = 1$), the definition becomes $\Delta\Pi_1 = \sqrt{2}\pi^2\left(\int_{r_1}^{r_2} N\frac{dr}{r}\right)^{-1}$, which, requiring only the observation of the most readily available modes, is the most reliable and commonly used diagnostic quantity for $g$-modes.

These two parameters, relating the outside shape and internal structure, can be used to describe the evolution of stars in a way similar to luminosity and effective temperature, the quantities used in the Hertzsprung–Russell (H-R) diagram.

Measurements show a marked distinction between stars in the ascending RGB and the red clump (RC), with the latter exhibiting much higher values of $\Delta\Pi_1$ (e.g., Hekker & Christensen-Dalsgaard 2017). However, the gap, spanning from one stage of evolution to the other, contains significantly fewer observations than the predicted band and cluster where stars in these stages respectively accumulate. Previous work by Bildsten et al. (2012) identified this region, allowing for the characterization of stars as undergoing the helium flash, and some candidates have been observed (Mosser et al. 2014; Vrard et al. 2016). A representation of these data superimposed with the asteroseismic evolution tracks of some selected models can be seen in Figure 1.

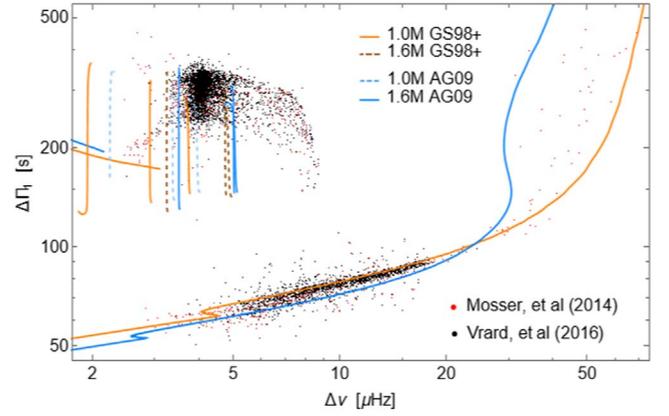

**Figure 1.** Asteroseismic evolution diagram from TAMS to the RC showing selected tracks superimposed on observational data taken from Mosser et al. (2014) and Vrard et al. (2016).

The region around $\Delta\nu \approx 4\,\mu$Hz and $\Delta\Pi_1 \approx 300$ s mostly corresponds to stars in the RC. Stars progress from the upper right of the plot to lower values of $\Delta\Pi_1$ as the size of the cavity where $g$-modes propagate increases, as more helium accumulates in the core and the fusion shell is pushed outward, in accordance with Equation (2). Additionally, due to the presence and properties of degenerate helium, the increasing similarity between the cores of RGB stars leads to a narrowing of the allowed values of $\Delta\Pi_1$, leading to the band feature at $50\,\text{s} < \Delta\Pi_1 < 100\,\text{s}$, while the continued expansion of the outer layers of the star, increasing its radius, contributes to the decrease of $\Delta\nu$.

According to Equation (2), the appearance of convection in the core (see, e.g., the third panel in Figures 2(a) and 2(b)) significantly reduces the size of the $g$-mode cavity, leading to the sharp increases in $\Delta\Pi_1$ (the near-vertical lines in Figure 1). The most significant increase happens during the helium flash (not shown), but the extension of the flash-driven convection zone is too large and happens too quickly to allow for observation of relevant modes.

However, if the stellar models prove accurate, the existence of a cooling mechanism in the inert core (via thermal neutrino emission, further explored in Section 4) causes the helium flash to occur in several parts. After the main, larger flash, several short-lived subflashes occur, generating convection in smaller regions.

In the asteroseismic evolution plot in Figure 1, the colored vertical lines note the predicted regions where the indicated models would undergo these subflashes, in a region spanning $100\,\text{s} < \Delta\Pi_1 < 300\,\text{s}$ in the $\Delta\nu - \Delta\Pi_1$ parameter space, part of which ($100\,\text{s} < \Delta\Pi_1 < 200\,\text{s}$) is unpopulated. For these models, the predicted values of the global asteroseismic parameters can be found in Table 3.

Note that, for the 1.0 $M_\odot$ models, the first subflash has a duration of approximately $10^4$ yr, and in that time, a variation of $\Delta\Pi_1$ of around 200 s is measured, of which slightly over half is spent in the unpopulated region of the parameter space. The stability of $\Delta\nu$ in this region could be of use in determining the composition of the observed stars. Our models predict differences of approximately 0.3 $\mu$Hz for stars with metallicity differences of 26.9%. This is in line with the operational expectations for the PLATO mission, which expects frequency uncertainties in the range of 0.14–0.28 $\mu$Hz for 48 months of data, assuming the uncertainty relation with time described in Goupil (2017). The level of precision after only

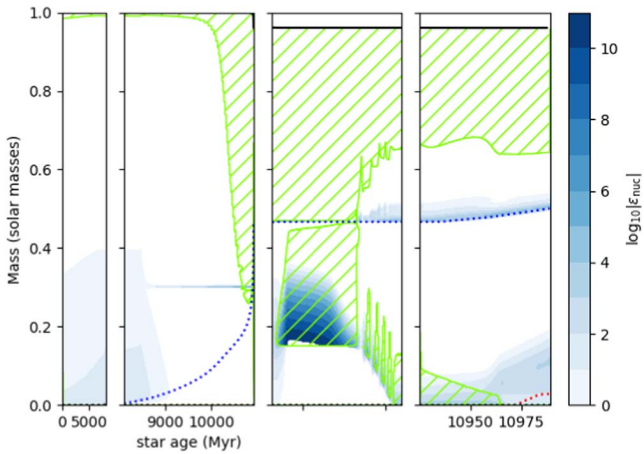

(a) 1.0 $M_\odot$

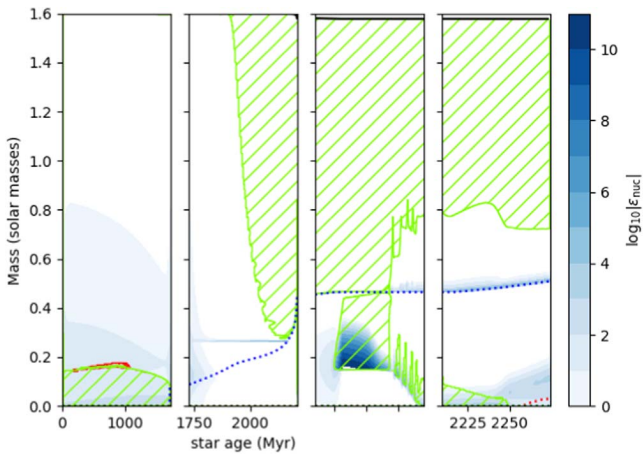

(b) 1.6 $M_\odot$

**Figure 2.** Kippenhahn diagrams for the 1.0 and 1.6 $M_\odot$ AG09+ models. The panels show, in order, the main sequence up to TAMS, TAMS to TRGB, helium flash, and postflash evolution until the formation of a carbon core (red dashed line). Green hatched areas denote convection. The dashed blue line delimits the helium core. Shaded blue areas indicate the energy generation rate of nuclear reactions ($\varepsilon_{\rm nuc}$) in units of erg g$^{-1}$ s$^{-1}$. The age axis is not linear in the helium flash panel.

**Table 3**
Asteroseismic Description of Points from the TAMS to the Third Core Subflash (SF) for the 1.0 $M_\odot$ GS98− Model

| Point | Age (Myr) | $\Delta\Pi_1$ (s) | $\Delta\nu$ ($\mu$Hz) | $R$ ($R_\odot$) |
|---|---|---|---|---|
| TAMS | 8326.293 | 0 | 97.19 | 1.294 |
| TRGB | 11,058.09 | 28.41 | 0.093 | 154.1 |
| N-flash | 11,058.10 | 104.4 | 0.095 | 151.6 |
| SF1 (start) | 11,058.29 | 128.3 | 1.959 | 17.92 |
| SF1 (end) | 11,058.31 | 363.0 | 2.040 | 17.44 |
| SF2 (start) | 11,058.53 | 138.3 | 3.022 | 13.38 |
| SF2 (end) | 11,058.55 | 317.1 | 3.000 | 13.45 |
| SF3 (start) | 11,058.84 | 145.0 | 3.921 | 11.24 |
| SF3 (end) | 11,058.86 | 281.5 | 3.829 | 11.42 |

**Note.** The points denoted SF correspond to the core subflashes in chronological order.

24 months is expected to be of the order of 0.2–0.4 $\mu$Hz for several modes around $\nu_{\rm max}$, which in itself could also be enough to distinguish stars by their compositions, assuming other methods are available to constrain their mass and radius.

### 4. Neutrinos

Though neutrino astronomy is still in its infancy, neutrinos make very appealing stellar probes, since they are freely emitted from stars at all stages of evolution, carrying with them information about the conditions in which they were created and arriving at detectors without interference. However, their lack of interactivity makes them notoriously hard to detect as well, with experiments being difficult to set up and modify and acquisition of sufficient data taking a long time.

We refer to nuclear neutrinos as those that are produced as the by-product of nuclear reactions (by decay of nuclear reaction products) and thermal neutrinos, which encompass all other sources of neutrino production in a star.

#### 4.1. Thermal Neutrinos

Other than being produced in fusion reactions or as products of the decay of nuclei involved in fusion, neutrinos may also originate from thermal processes (e.g., Compton scattering, pair annihilation, brehmsstrahlung, plasmon decay; Paxton et al. 2011) in a star. Although their existence is well established by theory, neutrinos from these sources have never been observed, and models of their behavior must rely exclusively on theoretical predictions (Itoh et al. 1996).

By the TRGB, the helium core is almost isothermal, and the only process relevant to the regulation of its temperature is, in fact, the emission of neutrinos via plasmon decay ($\gamma_{\rm plasmon} \longrightarrow \nu_e + \bar{\nu}_e$; Salaris et al. 2002). The total emission depends on the mass of the core and so increases steadily throughout the RGB phase. The existence of a cooling process causes a local inversion of the temperature gradient that is more pronounced at the points in the star's radius where neutrino emission is greatest. Traditionally, however, in the study of low-mass stars, thermal neutrinos are disregarded, since they contribute to the total neutrino emission by around 2 orders of magnitude less than the nuclear neutrinos during almost the entire evolution of the star up to the end of the RGB. The Kippenhahn diagrams in Figure 3 show the differences in neutrino emission via nuclear and thermal processes during the helium flash. Note that, although usually contributing around 100 times less to the total emission amount, thermal processes dominate inside the preflash, degenerate helium core, where no nuclear reactions take place.

Thermal neutrinos (which are, as we have noted, predominantly plasmon decay neutrinos) are, then, instrumental in the helium flash process, as their emission profile defines its starting point and, therefore, how the core is lifted out of degeneracy over time.

Helium flash observations may then provide an indirect way of testing thermal neutrino models. Their direct relation to the depth at which the helium flash starts and, therefore, the internal structure of the star during the subflashes means that analysis of the interior of the star (possible via asteroseismology in some cases, as noted by Miller Bertolami et al. 2020) should be able to test our understanding of these processes and their underlying assumptions. To that end, we compared the evolution of our 1.6 $M_\odot$ GS98+ model (which considers plasmon decay neutrino emission), in terms of the asteroseismic parameters $\Delta\nu$ and $\Delta\Pi_1$, with a version of the same model in which the only change was to completely suppress plasmon decay neutrino emission.

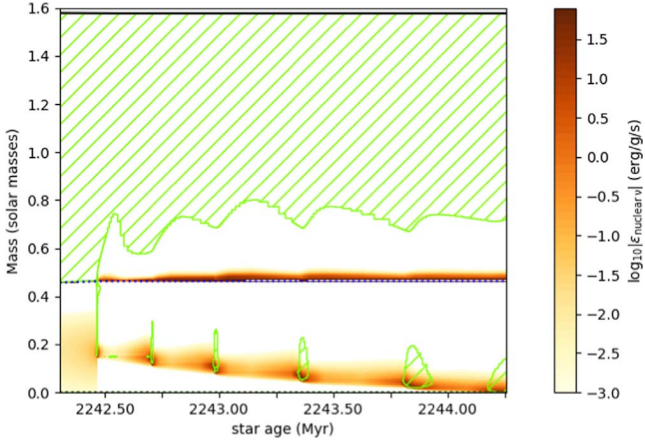

(a) nuclear neutrinos

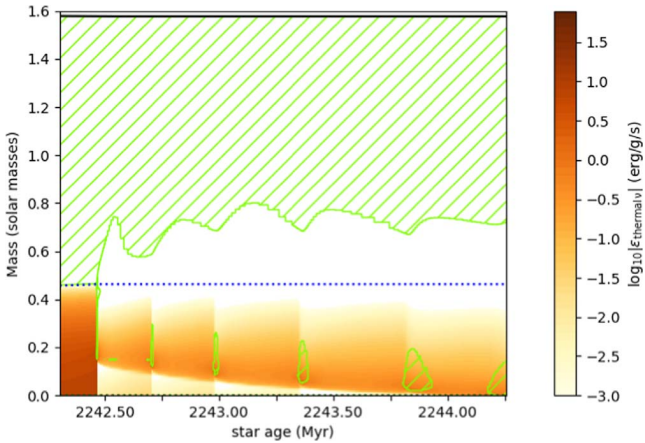

(b) thermal neutrinos

**Figure 3.** Internal structure and neutrino emission during the helium flash and subflashes for the 1.6 $M_\odot$ GS98+ model. Green hatched areas denote convection. The dashed blue line delimits the helium core. Shaded orange areas indicate the energy generation rate of neutrinos from nuclear sources, $\varepsilon_{\text{nuclear }\nu}$ (top), and thermal sources, $\varepsilon_{\text{thermal }\nu}$, predominantly plasmon decay (bottom), in units of erg g$^{-1}$ s$^{-1}$.

The asteroseismic diagnostic in Figure 4 reveals different structures for the post–helium flash core when the emission of plasmon decay neutrinos is and is not allowed, with the preflash tracks of the stars being identical. The model in which plasmon decay neutrinos were suppressed performs the helium flash in a single event and presents a comparatively smooth transition, with no subsequent subflashes being necessary, since the convection caused by the deflagration of helium burning at the stellar center lifts the entire core out of degeneracy. Asteroseismology is sensitive to the impact of this transition on, for instance, the location of the base of the (upper, larger) convective zone, illustrated for the case in which plasmon decay neutrinos are suppressed in the bottom panel of Figure 5. In contrast, the top panel reveals that each period of transient convection from a subflash pushes the base of the upper convective zone away from the stellar center, creating a less constant evolution pattern. The electron degeneracy factor, $\eta$,

$$\eta = \frac{\mu}{k_B T}, \quad (3)$$

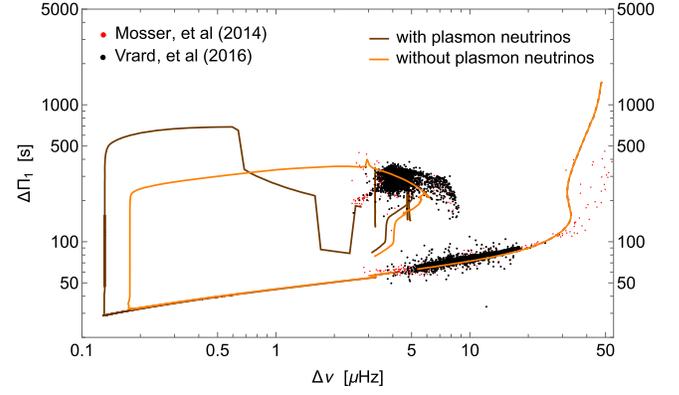

**Figure 4.** Comparison of asteroseismic evolution tracks for the 1.6 $M_\odot$ GS98+ model and the same model with plasmon decay neutrinos completely suppressed, superimposed on observational data taken from Mosser et al. (2014) and Vrard et al. (2016).

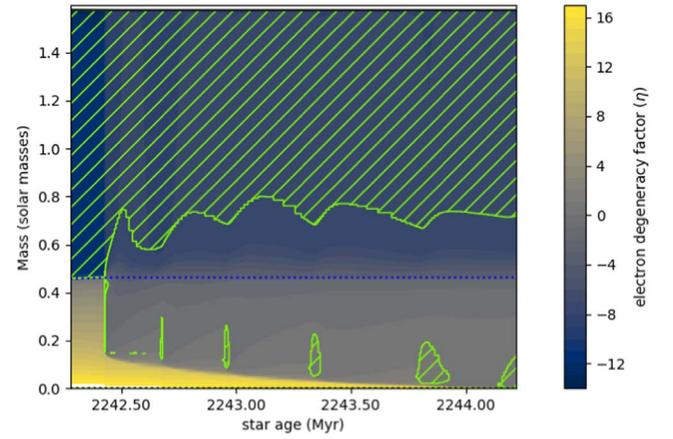

(a) plasmon decay neutrinos allowed

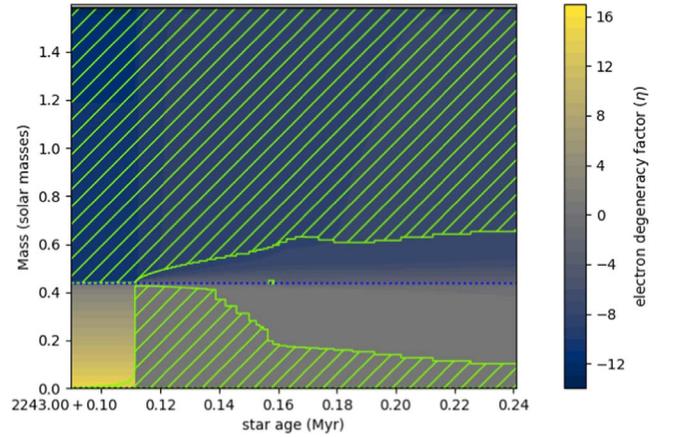

(b) plasmon decay neutrinos suppressed

**Figure 5.** Internal structure during the helium flash for the GS98+ 1.6 $M_\odot$ model with plasmon decay neutrinos enabled (top) and disabled (bottom). Green hatched areas denote convection. The dashed blue line delimits the helium core. Background areas are colored in a gradient describing the electron degeneracy factor, $\eta$, of Equation (3).

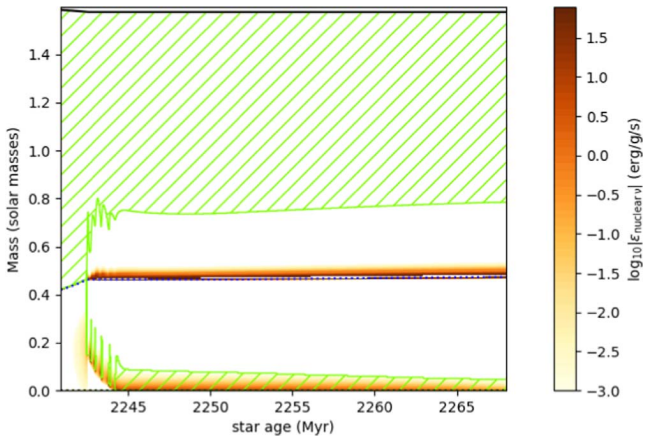

(a) plasmon decay neutrinos allowed

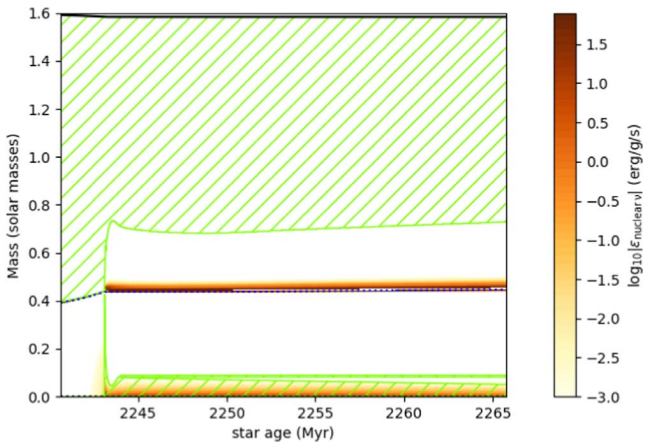

(b) plasmon decay neutrinos suppressed

**Figure 6.** Internal structure and neutrino emission during and for some time after the helium flash for the 1.6 $M_\odot$ GS98+ model. Green hatched areas denote convection. The dashed blue line delimits the helium core. Shaded orange areas indicate the energy generation rate of neutrinos from nuclear sources, $\varepsilon_{\mathrm{nuclear}\,\nu}$, in units of erg g$^{-1}$ s$^{-1}$.

where $\mu$ is the electron chemical potential, $k_B$ is the Boltzmann constant, and $T$ is the temperature, provides an indication of the degree of degeneracy of matter locally, with $\eta \gg 1$ indicating strongly degenerate matter. The regular model, with plasmon decay neutrinos enabled, gradually lifts the core out of degeneracy, as smaller flash events are triggered in the regions where helium remains accumulated, whereas in the model with plasmon decay neutrinos suppressed, the core transitions out of degeneracy immediately at the main (and only) flash event.

The longer-term structure of the star, however, is not affected by the way in which the core becomes nondegenerate (another characteristic that makes this process evade detection). Within a few million years, the cores settle into a pattern of helium burning, causing small inner convective areas (see Figure 6) in each case, and by the time convection in the core ends, the relative difference between the $\Delta\Pi_1$ of the models (as can be seen in the endpoints of the tracks in Figure 4) is 1.71%.

This configuration establishes a specific window of evolution where measurement of asteroseismic quantities can help constrain thermal neutrino theoretical predictions.

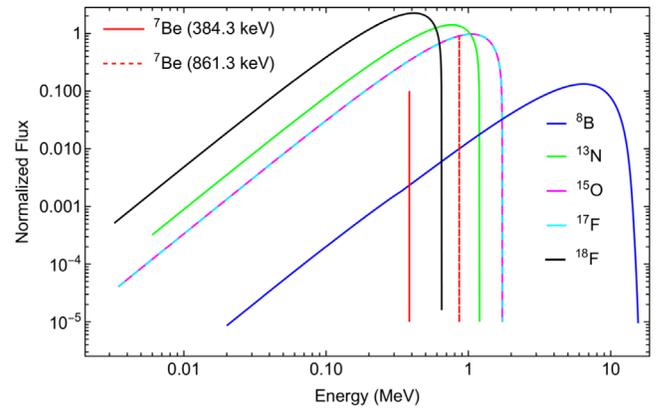

**Figure 7.** Neutrino energy spectrum for relevant pp and CNO cycle sources. The data were taken from Bahcall & Ulrich (1988).

### 4.2. Nuclear Neutrinos

The detection of (nuclear) solar neutrinos was instrumental in resolving the neutrino oscillation problem (Ahmad et al. 2001), a long-standing problem in astrophysics. More recently, the first neutrinos from the CNO reaction occurring in the Sun were measured in the Borexino experiment (Borexino Collaboration 2020), confirming models for fusion inside the Sun and providing the first clues toward a solution of the abundance problem. Though progress is being made, detecting the predominantly low-energy neutrinos emitted in the course of nuclear fusion from extrasolar sources remains out of reach to this day, as it would take extreme sources of emission that are relatively nearby for detection to be possible. This type of direct detection is, at this time, possible only for very specific sources and types of emission, namely, Type II supernovae (Hirata et al. 1988) and active galactic nuclei (IceCube Collaboration et al. 2022). Nonetheless, precise predictions of neutrino fluxes are valuable for comprehending the evolution of these stars and can also be beneficial for other astrophysical investigations, and the helium flash is a typical event found in low-mass stars with a considerable neutrino emission that can be accessed through asteroseismology.

Among the processes that are initiated by the rapid ignition of helium in the core is the burning of $^{14}$N into $^{18}$F ($^{14}$N + $\alpha \rightarrow ^{18}$F + $\gamma$), which then decays into $^{18}$O, emitting a neutrino ($^{18}$F $\rightarrow ^{18}$O + $e^+$ + $\nu_e$). These neutrinos are also low-energy, with an endpoint of 0.633 MeV and an average energy of 0.382 MeV. In this sense, they are much closer to the CNO cycle neutrinos than the more energetic $^8$B ones ($E < 15$ MeV) targeted by earlier experiments (see Figure 7). Despite these difficulties, the previously highlighted results from Borexino showcase the capability of detecting low-energy neutrinos of solar provenance.

The equilibrium mechanics of the CNO cycle by the end of the RGB mean that a large quantity of $^{14}$N is accumulated in the core, composing approximately 70% of all metal content there. When helium finally ignites, the conditions in the core cause nitrogen to deflagrate as well, being quickly consumed in what can be called the nitrogen flash (N-flash; Serenelli & Fukugita 2005). Though the prevalent $3\alpha$ process that turns $^4$He into $^{12}$C is not relevant to neutrino emission, it triggers the production of $^{18}$F from $^{14}$N, and the $^{18}$F decay neutrinos could serve as an effective signature of the helium flash, acting as a tool to observe this phenomenon directly for the first time. Additionally, since the flux of neutrinos from this reaction depends on the abundance of $^{14}$N in the stellar core, once detectable, these neutrinos could serve as a probe of element composition, helping to complete the panorama of stellar abundance for different regions of a star. This is especially worthy of note because CNO neutrinos are also a probe of $^{13}$N abundance in the fusion-active stellar region. Our predictions of the emissions that could be expected in our models may serve to constrain the expected impact that helium flash events would have on present-day experiments.

A picture of the evolution of the neutrino flux for each of the main nuclear sources ($^7$Be, $^8$B, and $^{13}$N) and the $^{18}$F N-flash neutrinos can be seen in Figure 8 for the 1.3 $M_\odot$ models. The full plots for all masses can be found in Appendix B. In each case, the abrupt, extremely quick decrease in flux for the main nuclear sources can be seen at the start of the helium flash, followed a short time later (at the nitrogen flash) by an increase in the flux of $^{18}$F neutrinos. It is also worthy of note how the preflash neutrino emission diverges little across all models and masses, a consequence of late RGB evolution being driven by the fusion layers lying on top of very similar degenerate cores.

The fluxes at the terminal-age main sequence (TAMS) and the nitrogen flash for a selection of models are available in Tables 4 and 5. Note how the neutrino flux pertaining to the reactions of the different pp branches ($^7$Be for pp branch II and $^8$B for pp branch III) and the CNO cycle ($^{13}$N) change their predominance across the different mass values.

The expected magnitude of the emission of $^{18}$F neutrinos during the nitrogen flash can be seen to be, for all masses and compositions, 10 orders of magnitude above the expected emission of the TAMS CNO neutrinos. A helium flash event of a 1 $M_\odot$ star would be measured on Earth as 10% of the current-day flux of $^{13}$N neutrinos from the Sun (Vinyoles et al. 2017) if they were between approximately 23 (for the GS98 models) and 25 (for the A09 models) pc away.

Additionally, it is relevant to note that the $^{18}$F neutrinos are the only ones whose production is expected to vary linearly with metallicity for any mass. The circumstances of their production, being dominated by the onset of the helium flash (which depends only on the core mass and thermal neutrinos), causes them to be unaffected by the more indirect effects of changing metallicity and mass, namely, the changes in the relative dominance of hydrogen-burning nuclear reactions. This, particularly when paired with other independent methods of determining the mass of the star, could be of great use in discerning metallicity, especially when comparing similar objects, and measurements of this neutrino flux could help establish a metallicity map of observed (distant) sources.

Our models also allow us to analyze the profile of neutrino emissions by source, allowing us to pinpoint the location where neutrinos are being produced. In Figure 9, an example plot of the full data can be seen for the 1.6 $M_\odot$ GS98+ model. Comparison with Figures 12–15 (in Appendix B) reveals that, despite the significant difference in expected fluxes, the location in the star where the neutrinos are being produced varies little from the helium to the nitrogen flash, as would be expected because only a short time elapses between them.

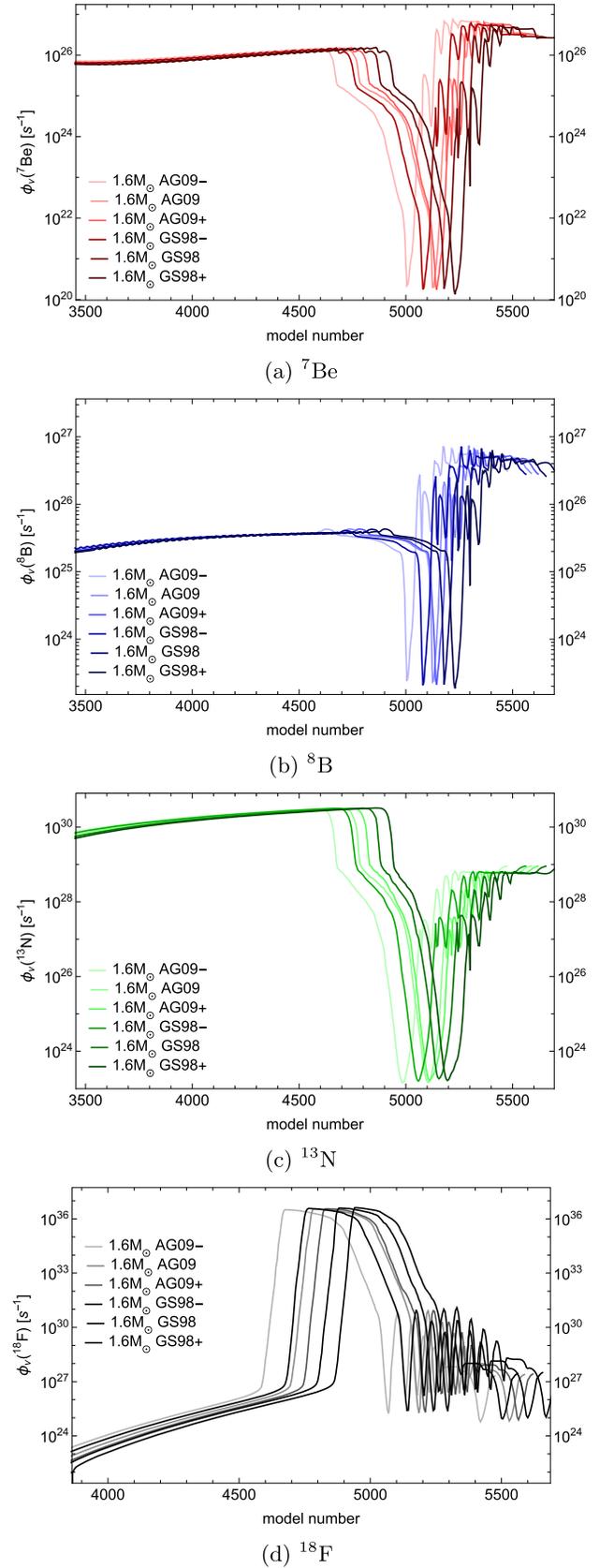

**Figure 8.** Neutrino flux (in s$^{-1}$) over time for the considered nuclear sources for models of 1.6 $M_\odot$. The plot is not linear on the time axis.

**Table 4**
Neutrino Flux per Source of Emission at TAMS for the GS98+, GS98−, AG09+, and AG09− Compositions and All Masses

| Flux (s$^{-1}$) | 1.0 $M_\odot$ | | | |
|---|---|---|---|---|
| | GS98+ | GS98− | AG09+ | AG09− |
| $^7$Be($\times 10^{26}$) | 6.91 | 7.86 | 7.73 | 8.72 |
| $^8$B($\times 10^{24}$) | 2.97 | 3.89 | 3.74 | 4.84 |
| $^{13}$N($\times 10^{26}$) | 6.39 | 6.93 | 6.90 | 7.45 |
| $^{18}$F | 0 | 0 | 0 | 0 |
| | 1.3 $M_\odot$ | | | |
| $^7$Be($\times 10^{26}$) | 16.5 | 18.7 | 19.2 | 21.5 |
| $^8$B($\times 10^{24}$) | 34.5 | 45.1 | 47.3 | 60.1 |
| $^{13}$N($\times 10^{26}$) | 30.3 | 33.4 | 32.9 | 35.2 |
| $^{18}$F | 0 | 0 | 0 | 0 |
| | 1.6 $M_\odot$ | | | |
| $^7$Be($\times 10^{26}$) | 16.5 | 18.7 | 19.6 | 22.0 |
| $^8$B($\times 10^{24}$) | 54.6 | 71.9 | 78.8 | 102 |
| $^{13}$N($\times 10^{26}$) | 116 | 128 | 126 | 135 |
| $^{18}$F | 0 | 0 | 0 | 0 |

**Table 5**
Neutrino Flux per Source of Emission at the Nitrogen Flash for the GS98+, GS98−, AG09+, and AG09− Compositions and All Masses

| Flux (s$^{-1}$) | 1.0 $M_\odot$ | | | |
|---|---|---|---|---|
| | GS98+ | GS98− | AG09+ | AG09− |
| $^7$Be($\times 10^{26}$) | 0.53 | 0.50 | 0.50 | 0.48 |
| $^8$B($\times 10^{24}$) | 20.6 | 20.3 | 20.7 | 21.1 |
| $^{13}$N($\times 10^{26}$) | 80.6 | 85.0 | 92.0 | 103 |
| $^{18}$F($\times 10^{36}$) | 4.77 | 4.62 | 3.98 | 3.47 |
| | 1.3 $M_\odot$ | | | |
| $^7$Be($\times 10^{26}$) | 0.44 | 0.41 | 0.41 | 0.42 |
| $^8$B($\times 10^{24}$) | 22.2 | 19.9 | 17.4 | 20.4 |
| $^{13}$N($\times 10^{26}$) | 1070 | 977 | 744 | 926 |
| $^{18}$F($\times 10^{36}$) | 4.29 | 3.94 | 3.85 | 2.99 |
| | 1.6 $M_\odot$ | | | |
| $^7$Be($\times 10^{26}$) | 0.35 | 0.34 | 0.35 | 0.36 |
| $^8$B($\times 10^{24}$) | 17.5 | 19.1 | 19.8 | 21.9 |
| $^{13}$N($\times 10^{26}$) | 83.2 | 104 | 97.1 | 112 |
| $^{18}$F($\times 10^{36}$) | 4.20 | 3.79 | 3.61 | 3.21 |

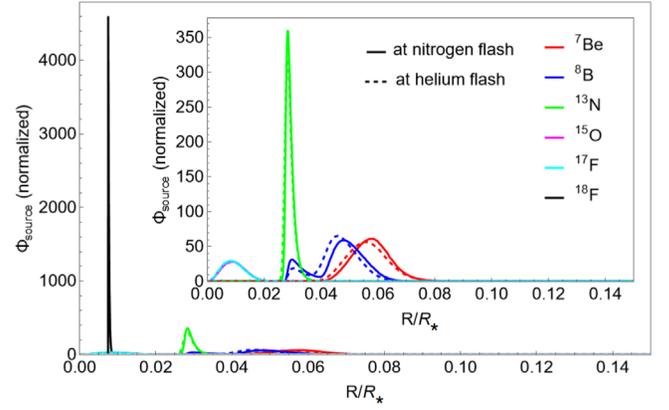

**Figure 9.** Comparison of neutrino emission profiles by source at the helium flash (dashed) and nitrogen flash (solid) for the 1.6 $M_\odot$ GS98+ model. The strongest source ($^{18}$F) is absent in the inset.

## 5. Conclusions

In this work, we have analyzed neutrino fluxes and their impact on the observable asteroseismology of 18 stellar models from the TAMS to the late RGB and through the helium and nitrogen flashes up to the formation of a carbon core across a range of masses (1.0 $M_\odot \leqslant M_* \leqslant$ 1.6 $M_\odot$), metallicities (0.0115 $\leqslant Z_* \leqslant$ 0.0183), and relative element abundances (GS98 and AG09). The focus was to determine if these parameters could serve as evidence of the predicted but never yet directly observed helium flash and the form in which it occurs and to ascertain the impact of changes in stellar composition in the evolution of neutrino fluxes, as well as on global asteroseismic parameters, in the context of the stellar abundance problem.

We demonstrate that thermal neutrinos play a determining role in the shape of the helium flash and that these differences are available for asteroseismic observation via the global asteroseismic parameters $\Delta\nu$ and $\Delta\Pi_1$. We also predict the neutrino fluxes separately for each source, including the potentially relevant $^{18}$F decay, which occurs as a product of the deflagration of helium and could serve (in the future) as evidence of this event. For this source, the expected emissions are on the order of $\approx$3–4 $\times 10^{36}$ s$^{-1}$ for stars of all masses and metallicities, and with current technology, a similar event would lead to a 10% perturbation of the neutrino flux from the Sun for stars within 25 pc of Earth.

Finally, we have determined that our models replicate well the available observational asteroseismic data for subgiant and RGB stars. Additionally, they confirm that the core subflashes that, according to stellar models, occur after the (first and) main helium flash cause the $\Delta\Pi_1$ of the stars to cross the expected approximate 100–280 s region in an otherwise empty portion of the asteroseismic evolution diagram. We find differences of 0.3 $\mu$Hz in the $\Delta\nu$ of the AG09 and GS98+ models, which should be resolved by the PLATO mission after a period of observation between 24 and 48 months.

## Acknowledgments

I.L. thanks the Fundação para a Ciência e Tecnologia (FCT), Portugal, for the financial support to the Center for Astrophysics and Gravitation (CENTRA/IST/ULisboa) through grant project No. UIDB/00099/2020 and grant No. PTDC/FIS-AST/28920/2017.

The authors would like to acknowledge the efforts of Bill Paxton and the other 19 current and past developers on the MESA team for providing accurate, easy-to-use, and powerful software tools for free and for documenting, maintaining, and constantly improving the code. They would also like to thank the MESA users mailing group members for their availability to answer users' questions.

They would also like to acknowledge the work of Pablo Marchant in the code that allows for the plotting of the Kippenhahn diagrams from MESA data.

Finally, the authors would like to thank the anonymous referee for the comments and suggestions regarding the focus of the results.

This work has made use of NASA's Astrophysics Data System Bibliographic Services.

## Data Availability

The MESA stellar evolutionary code is available for free in the public domain and can be found at the MESA Zenodo repository (Paxton 2019).

The observational data we used are readily and openly available and can be found in the VizieR catalogs (Mosser et al. 2014; Mosser et al. 2014) and Vrard et al. (2016).

All of the inlists and custom files used with MESA to obtain these results are made available, according to best practices, at the Zenodo repository with the doi:10.5281/zenodo.7836792.

## Appendix A
## Tip of the Red Giant Branch

To establish a quality benchmark for the evolved models used in this work, we performed an analysis of the stability of the TRGB. The general evolution of the models can be seen in the H-R diagram in Figure 10.

A detail of the region around the TRGB is presented in Figure 11 for all models.

The regularity of the luminosity of stars as they reach the TRGB can be seen as a consequence of the similarity of their cores. While this may be of use to astronomers as a standard candle, its lack of variation makes it difficult to identify a star's fundamental properties from its value. As can be seen in Table 6, an increase of 0.3 $M_\odot$ in the models results in a difference of around 1% in luminosity. The $T_{eff}$, while exceeding the 2% average uncertainty of determinations (Tayar et al. 2022; Saltas et al. 2022) between stars with the exact same composition and different mass, does not correlate so directly if one allows the stars' metallicity to vary (see Figure 11).

Relatively small differences in metallicity and composition do not change the processes through which the star evolves; rather, they determine how quickly the star will move from one phase to the next. Stars with a higher metallicity present a higher opacity, and differences in relative metal abundances may shift the picture of equilibrium in certain reactions, like the CNO cycle.

Due to their relative ease of determination from astronomical observations and the characteristics of stellar evolution (which lead stars with fundamentally different properties to exhibit different sets of these two quantities), effective temperature ($T_{eff}$) and bolometric luminosity ($L$) can often be used to ascertain the evolutionary stage of a star and put bounds on its other properties. However, this analysis is not without limits, especially when the evolutionary track of a star crosses over a region of the H-R diagram parameter space several times. Considering specifically red giants pre– and post–helium flash, these stars are very different structurally but not very different when looked at broadly from the outside (as previously mentioned, this is one of the difficulties in directly detecting a helium flash event). As Figure 11 shows, uncertainties in metallicity and composition can make it difficult to determine a star's fundamental quantities based solely on its position in the H-R diagram, since stellar metallicities can seldom be determined with a precision greater than the one assumed between the models shown. Additionally, differences in relative element abundance between similar stars tend to manifest more expressively in their age (Capelo & Lopes 2020), a quantity that is not directly determinable from observation and relies on modeling or specific circumstances (e.g., the star being part of a cluster against whose background it can be compared) to be ascertained. For those reasons, it carries one of the largest uncertainties in determination from grid comparison (Tayar et al. 2022). Conversely, this makes determining a star's composition from luminosity and temperature (or other more accessible quantities) a difficult task. Table 7 exemplifies this point for stars of similar mass and metallicity but very different relative element abundances.

A summary of all TRGB luminosities for the modeled stars can be found in Table 8. The results show a maximum difference of less than 4% variation with respect to the mean luminosity across all models for the entire 1.0–1.6 $M_\odot$ range, with half the models differing from the mean by less than 1.2%. Even in the era of the phenomenal measurements made possible by Gaia and TESS, it is uncommon for determinations of luminosity to carry uncertainties below 1% (Heiter et al. 2015; which is still below the 1.6% theoretical uncertainty of current-generation, specific-purpose grid models), which would make the luminosities of roughly half of the stars in our model set indistinguishable from each other.

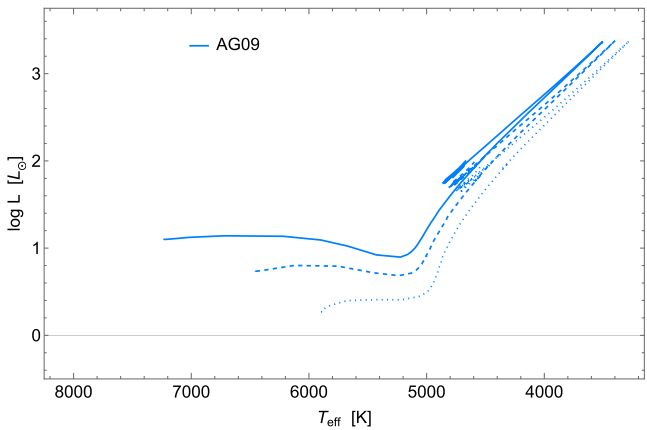

**Figure 10.** The H-R diagram showing stellar evolution tracks from the end of the main sequence (TAMS) to the RC for models of 1.0 (dotted), 1.3 (dashed), and 1.6 (solid) $M_\odot$ with the same metallicity and element abundances—in this case, the AG09+ model. Note the stability of the TRGB's luminosity across all models in the upper right corner.

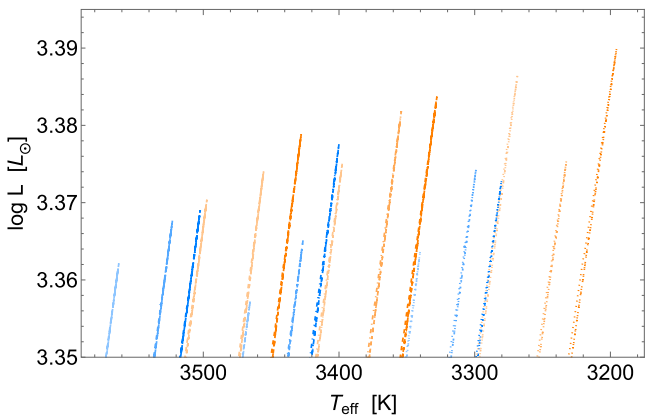

**Figure 11.** Zoom of the H-R diagram showing the TRGB for all 18 models. The models presented are 1.0 (dotted), 1.3 (dotted–dashed), and 1.6 (dashed) $M_\odot$ in blue for AG09 and orange for GS98 relative element abundances. Lighter and darker lines represent models with lower and higher metallicities, respectively.

**Table 6**
Summary of Stellar Quantities at the TRGB for Models of Equal Metallicity and Relative Element Abundance but Different Mass

| Mass ($M_\odot$) | 1.0 | 1.3 | 1.6 |
|---|---|---|---|
| Model | AG09+ | AG09+ | AG09+ |
| Age ($\times 10^6$ yr) | 11,665.1 | 4484.21 | 2207.96 |
| $T_{\rm eff}$ (K) | 3280 | 3400 | 3503 |
| $L$ ($L_\odot$) | 2362 | 2386 | 2334 |
| $T_c$ ($\times 10^7$ K) | 7.63 | 7.61 | 7.65 |
| $\rho_c$ ($\times 10^5$ g cm$^{-3}$) | 8.535 | 8.526 | 8.462 |
| $T_{\rm max}$ ($\times 10^7$ K) | 9.21 | 9.43 | 9.16 |
| $R_{T_{\rm max}}$ ($\times 10^{-3}$ $R_\odot$) | 7.82 | 7.81 | 7.69 |

**Note.** The subscript $c$ denotes a quantity at the stellar center.

**Table 7**
Summary of Stellar Quantities at the TRGB for Models of the Same Mass and Similar Metallicities but Different Relative Element Abundances

| Mass ($M_\odot$) | 1.6 | 1.6 | $\frac{|({\rm GS98}-) - ({\rm AG09}+)|}{({\rm AG09}+)}$ (%) |
|---|---|---|---|
| Model | GS98− | AG09+ | |
| Metallicity | 1.4650 | 1.4451 | |
| Age ($\times 10^6$ yr) | 2110.17 | 2207.96 | 4.43 |
| $T_{\rm eff}$ (K) | 3498 | 3503 | 0.14 |
| $L$ ($L_\odot$) | 2341 | 2334 | 0.33 |
| $T_c$ ($\times 10^7$ K) | 7.656 | 7.654 | 0.03 |
| $\rho_c$ ($\times 10^5$ g cm$^{-3}$) | 8.450 | 8.462 | 0.18 |
| $T_{\rm max}$ ($\times 10^7$ K) | 9.152 | 9.162 | 0.11 |
| $R_{T_{\rm max}}$ ($\times 10^{-3}$ $R_\odot$) | 7.75 | 7.69 | 0.78 |

**Table 8**
TRGB Luminosities for All 18 Models

| Mass ($M_\odot$) | Model | $L$ ($L_\odot$) | $\sigma_{\rm mass}$ (%) | $\sigma_{\rm all}$ (%) |
|---|---|---|---|---|
| 1.0 | AG09− | 2310.89 | 3.06 | 2.27 |
| | AG09 | 2367.95 | 0.66 | 0.15 |
| | AG09+ | 2359.12 | 1.03 | 0.23 |
| | GS98− | 2435.86 | 2.19 | 3.02 |
| | GS98 | 2373.68 | 0.42 | 0.39 |
| | GS98+ | 2454.78 | 2.98 | 3.82 |
| 1.3 | AG09− | 2276.05 | 3.70 | 3.74 |
| | AG09 | 2318.38 | 1.91 | 1.95 |
| | AG09+ | 2385.71 | 0.94 | 0.90 |
| | GS98− | 2372.48 | 0.38 | 0.34 |
| | GS98 | 2408.98 | 1.92 | 1.88 |
| | GS98+ | 2419.46 | 2.37 | 2.32 |
| 1.6 | AG09− | 2302.14 | 1.88 | 2.64 |
| | AG09 | 2332.11 | 0.61 | 1.37 |
| | AG09+ | 2339.04 | 0.31 | 1.08 |
| | GS98− | 2346.27 | 0.00 | 0.77 |
| | GS98 | 2366.01 | 0.84 | 0.06 |
| | GS98+ | 2392.45 | 1.97 | 1.18 |

**Note.** Here $\sigma_{\rm mass} = \frac{|L_{\rm Model} - \bar{L}_{\rm mass}|}{\bar{L}_{\rm mass}}$, where $\bar{L}_{\rm mass}$ represents the mean of the luminosities of models with the same mass, and $\sigma_{\rm all} = \frac{|L_{\rm Model} - \bar{L}|}{\bar{L}}$, where $\bar{L}$ represents the mean of the luminosities of all models.

# Appendix B
# Neutrino Emission Fluxes

The plots in Figures 12–15 show the evolution of neutrino flux per source over time for all of the models.

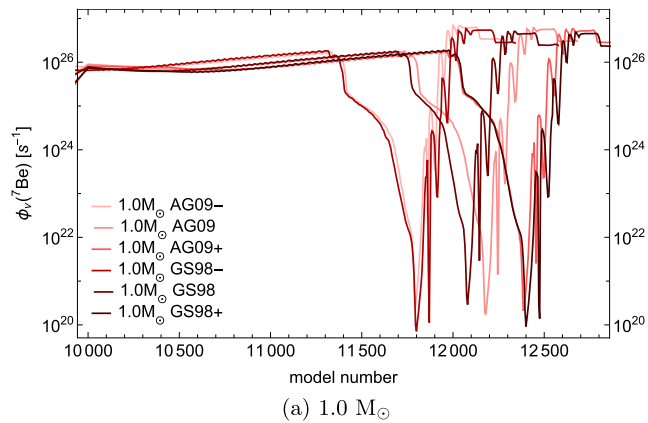

(a) 1.0 $M_\odot$

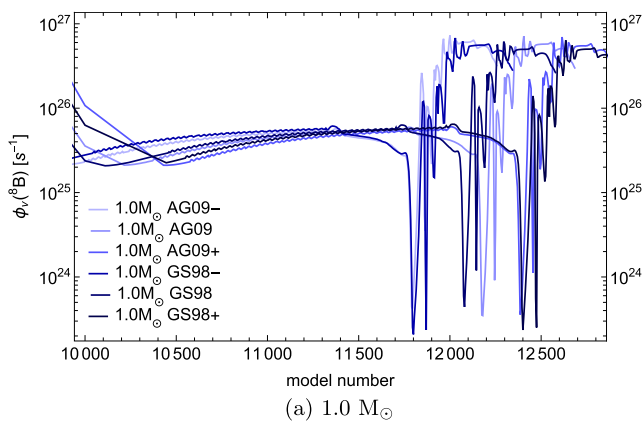

(a) 1.0 $M_\odot$

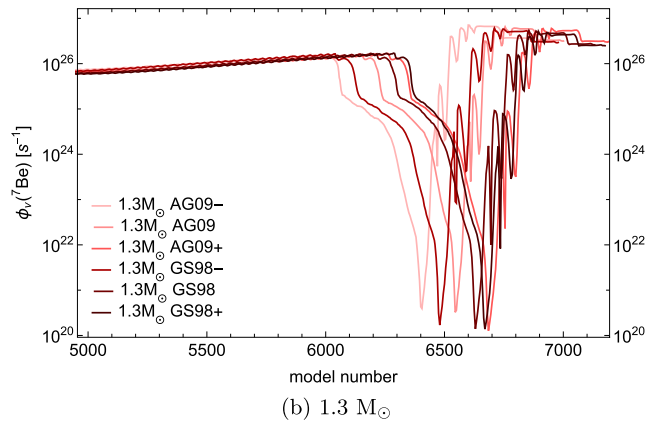

(b) 1.3 $M_\odot$

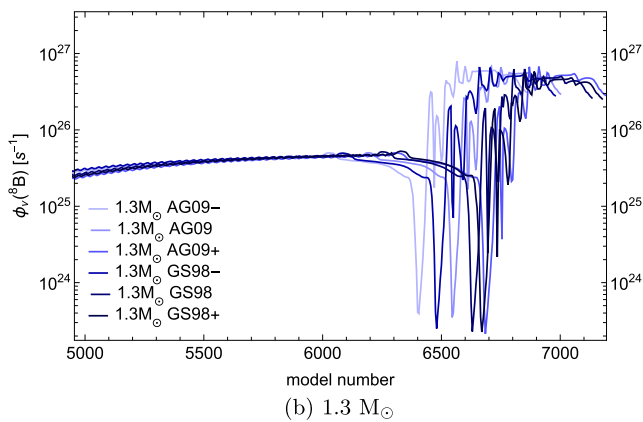

(b) 1.3 $M_\odot$

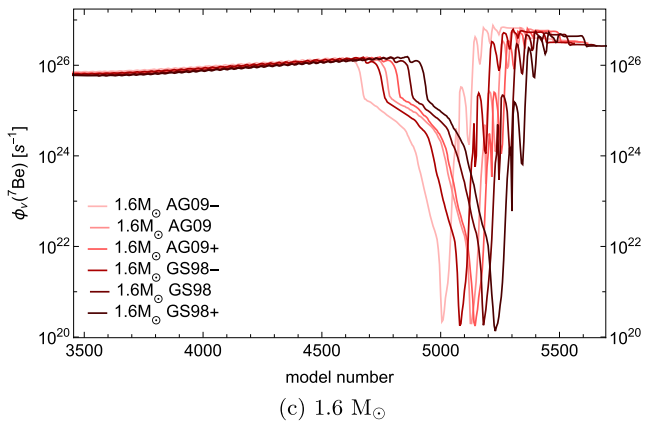

(c) 1.6 $M_\odot$

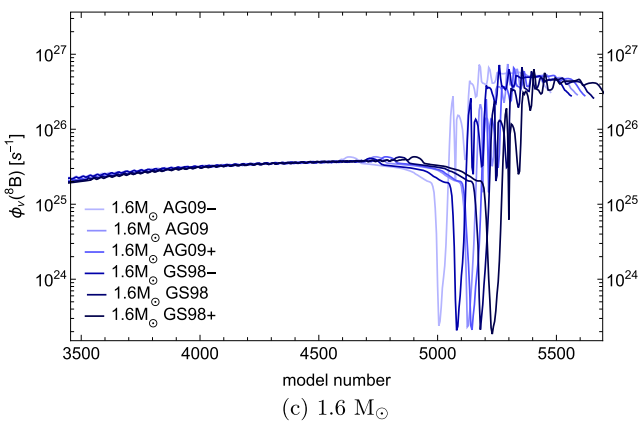

(c) 1.6 $M_\odot$

**Figure 12.** The $^7$Be neutrino flux (in s$^{-1}$) over time for all models, grouped by mass. The model number axis is not linear in time.

**Figure 13.** The $^8$B neutrino flux (in s$^{-1}$) over time for all models, grouped by mass. The model number axis is not linear in time.

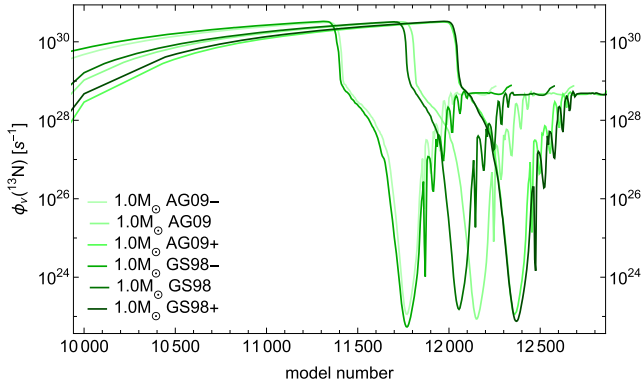
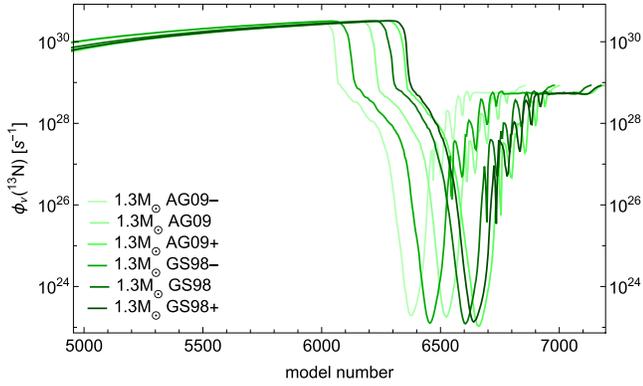
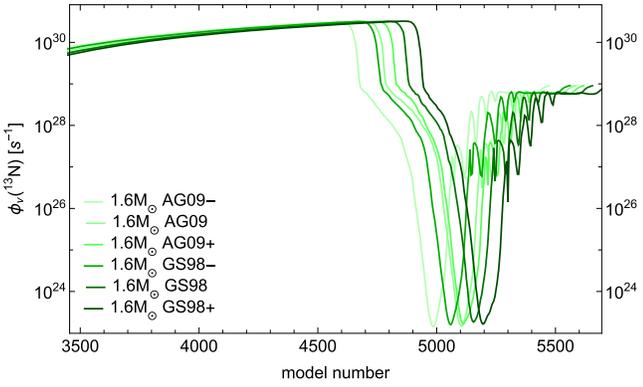

**Figure 14.** The $^{13}$N neutrino flux (in s$^{-1}$) over time for all models, grouped by mass. The model number axis is not linear in time.

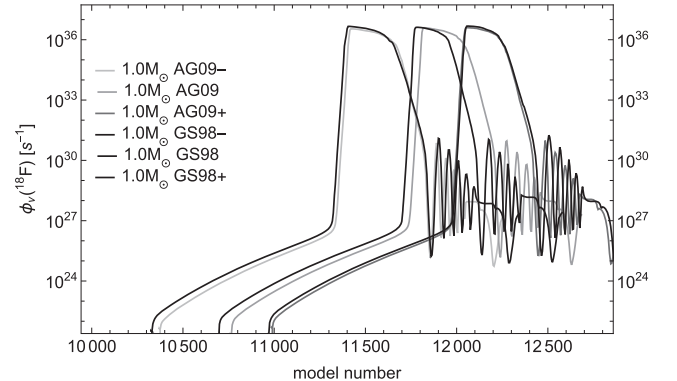
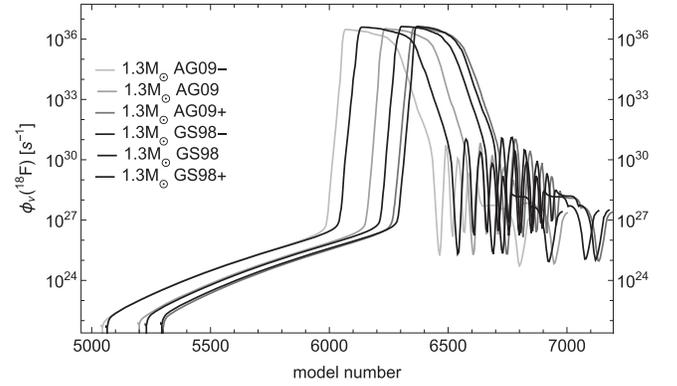
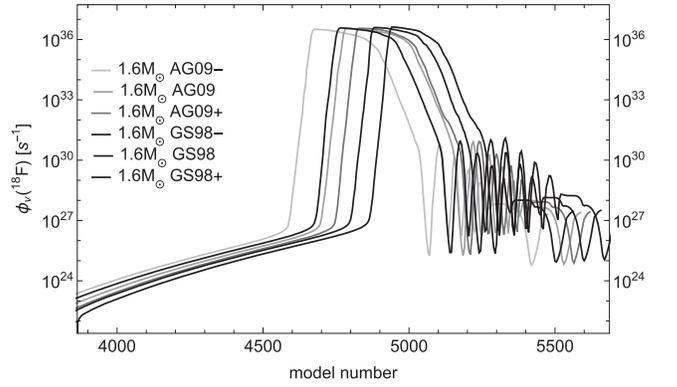

**Figure 15.** The $^{18}$F neutrino flux (in s$^{-1}$) over time for all models, grouped by mass. The model number axis is not linear in time.

## ORCID iDs

Diogo Capelo  https://orcid.org/0000-0002-3118-1284
Ilídio Lopes  https://orcid.org/0000-0002-5011-9195